\documentclass[aps,prd,onecolumn,groupedaddress,showpacs,nofootinbib,amssymb]{revtex4-2}
\usepackage[dvips]{graphicx}
\usepackage{amssymb}
\usepackage{amsmath}
\usepackage{graphicx,,color}
\usepackage{amsfonts}
\usepackage{bm}
\usepackage{cancel}
\usepackage{comment}
\usepackage{floatflt}

\newcommand\be{\begin{equation}}
\newcommand\ee{\end{equation}}

\allowdisplaybreaks[4]

\begin{document}

\title{A Remedy of the Trans-Planckian Censorship Problem with Smooth Slow-roll to Power-law Inflation Transitions in Scalar Field Theory}
\author{S.D. Odintsov$^{1,2,3}$}\email{odintsov@ice.csic.es}
\author{V.K. Oikonomou,$^{3,4}$}\email{voikonomou@gapps.auth.gr;v.k.oikonomou1979@gmail.com}
\affiliation{$^{1)}$ Institute of Space Sciences (ICE, CSIC) C. Can Magrans s/n, 08193 Barcelona, Spain \\
$^{2)}$ ICREA, Passeig Luis Companys, 23, 08010 Barcelona, Spain\\
$^{3)}$L.N. Gumilyov Eurasian National University - Astana,
010008, Kazakhstan\\
$^{4)}$Department of Physics, Aristotle University of
Thessaloniki, Thessaloniki 54124, Greece}


 \tolerance=5000

\begin{abstract}
It is known that if the standard slow-roll inflation is followed
by a power-law inflationary regime, then the trans-Planckian modes
may be safely be contained in the Hubble horizon and never exit it
during inflation. In this work we investigate how to realize a
smooth transition between a slow-roll and a power-law inflationary
regime in the context of single scalar field inflation. As we show
it is possible to realize such a smooth transition by generalizing
the kinetic energy of single scalar field in the form
$\dot{\phi}^2=\beta(\phi)V(\phi)$, where $\beta(\phi)$ is some
appropriate function of the scalar field. Using two distinct
approaches we show that it is possible to realize a smooth
transition from a slow-roll to a power-law inflationary regime,
and the two approaches produce identical results regarding the
slow-roll regime. Also we show that the slow-roll regime is quite
short, about $N\sim 30$ $e$-foldings, with the flatness and
horizon problems being solved with the synergistic effect of the
two inflationary patches. The slow-roll era is found to be
compatible with the Atacama Cosmology Telescope data.
\end{abstract}

\maketitle

\section{Introduction}

The inflationary era is one mysterious prediction of modern
theoretical cosmology regarding the early classical stages of our
Universe. This conceptual construction of modern theoretical
cosmology regarding the post-Planck primordial Universe, solves
all the consistency problems of the standard Big Bang scenario,
such as the flatness and horizon problems. There are two widely
accepted ways to describe the inflationary era, firstly, the
scalar field description
\cite{inflation1,inflation2,inflation3,inflation4}, which relies
heavily on the existence of a scalar field which controls the
early accelerating expansion of our Universe, while at the same
time having quite many couplings to the Standard Model particles
so that it can reheat the Universe after the inflationary era
ends. The scalar field description is motivated due to the fact
that spin zero bosons exist in nature, such as the Higgs boson,
and thus scalar fields may be remnants of the primordial
Ultraviolet Completion of the Standard Model of particles. The
second mainstream description is modified gravity in its various
forms \cite{reviews1,reviews2,reviews3,reviews4}, with the most
prominent candidate being $F(R)$ gravity, which is the simplest
extension of Einstein-Hilbert gravity. In both descriptions there
is the prediction of an early acceleration stage, with the
modified gravity being more appealing due to the fact that both
early and late-time acceleration eras can be described by the same
$F(R)$ gravity. This early acceleration can generate primordial
tensor and scalar perturbations with the latter being able to be
the source of large scale structure. In the following years
inflation will be further scrutinized and constrained by the
current and future experiments. The Planck collaboration is a
successor of previous telescopes, and has put quite stringent
constraints on the inflationary era \cite{Planck:2018jri}, without
however directly detecting signals that the inflationary era ever
existed. The smoking gun for the detection of inflation will be
the discovery of the $B$-mode in the Cosmic Microwave Background
(CMB) radiation. Recently the Atacama Cosmology Telescope (ACT)
has brought new data on the table, by constraining the scalar
spectral index of the primordial scalar curvature perturbations in
a different way compared to the Planck collaboration. This
discovery is somewhat interesting, although it is rather early to
consider such deviation from Planck quite seriously, however one
should also take this into account in order to be consistent
scientifically. New data from the Simons observatory
\cite{SimonsObservatory:2019qwx} or the future gravitational wave
experiments
\cite{Hild:2010id,Baker:2019nia,Smith:2019wny,Crowder:2005nr,Smith:2016jqs,Seto:2001qf,Kawamura:2020pcg,Bull:2018lat,LISACosmologyWorkingGroup:2022jok}
are expected to further shed light to the question on whether the
inflationary era has ever occurred. Regarding the gravitational
waves, the existence of a stochastic signal with small amplitudes
by LISA or the Einstein Telescope may originate from an
inflationary era. Such a stochastic background has been observed
in 2023
\cite{nanograv,Antoniadis:2023ott,Reardon:2023gzh,Xu:2023wog},
which however is highly unlikely to be described solely by an
inflationary era \cite{Vagnozzi:2023lwo,Oikonomou:2023qfz}.

Now regarding the inflationary era itself, although it is an
appealing theoretical construction, it comes along with its
thorns. The most serious conceptual problem is the possibility of
having trans-Planckian modes with very small wavelength, exiting
the Hubble horizon during inflation. This is quite possible, and
of course it is a source of inconsistency for inflationary
theories. The Planck epoch of our Universe is currently
unperceived, and although many theories, like string theory and
quantum gravity, provided some consistent framework for this
epoch, nobody really grasps this mysterious epoch. It belongs to
the quantum era of our Universe, where possibly all the
interactions are unified. Inflation on the other hand belongs to
the classical epoch of our Universe, thus the Universe is four
dimensional, and all the modes that exit the Hubble horizon during
inflation are frozen and evolve classically after first horizon
exit. Thus having trans-Planckian modes outside the Hubble horizon
is a rather unwanted feature. In order to solve this problem in
the context of a slow-rolling inflationary epoch, that is with a
quasi de Sitter evolution, one needs to put severe constraints on
the parameters of inflation, for example the scale of inflation
must be $V_e<10^{10}\,$GeV, and the tensor-to-scalar ratio must be
$r<10^{-30}$. This is quite restrictive and it is known as the
trans-Planckian Censorship Conjecture (TCC)
\cite{Martin:2000xs,Brandenberger:2000wr,Bedroya:2019snp,Brandenberger:2021pzy,Brandenberger:2022pqo,Kamali:2020drm,Berera:1995ie,Brandenberger:2025hof}.
In a previous work we presented a mechanism in the context of
$F(R,\phi)$ theory of gravity, which can provide a remedy in the
TCC problems of slow-roll inflation \cite{Odintsov:2025zrp}.
Specifically, inspired by Ref. \cite{Kamali:2020drm}, we showed
that if a standard $R^2$ slow-roll era is followed by a power-law
inflationary era, the TCC problems of slow-roll inflation can be
solved. In this work, we shall present a single scalar field model
which can realize a smooth transition from a slow-roll
inflationary epoch to a power-law inflationary epoch, which thus
can provide a remedy for the TCC problems of the single scalar
field slow-roll inflation. Our smooth transition will be achieved
by assuming that the scalar field evolves in the following way
$\dot{\phi}^2=\beta (\phi)V(\phi)$, where the function
$\beta(\phi)$ is chosen in such a way so that when the scalar
field values are large, the function $\beta(\phi)$ is
$\beta(\phi)\simeq 0$, so the slow-roll era is realized. On the
other hand, for smaller values of the scalar field, when the
slow-roll era ends, the function $\beta(\phi)$ behaves as
$\beta(\phi)=\beta$, where $\beta$ is some dimensionless constant,
and thus the power-law epoch can be realized. The results of our
analysis are quite interesting since, for choice of the function
$\beta(\phi)$ we used, the slow-roll era realized lasts for nearly
$N=30$ $e$-foldings, and is followed by a power-law inflationary
epoch, and thus the total duration of the inflationary era can
solve the flatness and horizon problems, especially if the
parameter $\beta$ is larger than $\beta>0.5$. One interesting
feature of this work is that the slow-roll era produces a scalar
spectral index which is compatible with the latest ACT data. We
studied the inflationary era using two patches, one slow-roll and
one power-law inflation patch, but we also used an alternative
formalism for the scenario that $\dot{\phi}^2=\beta (\phi)V(\phi)$
to study inflation. To this end we found expressions for the
inflationary indices and also the spectral index for scalar
perturbations and the tensor-to-scalar ratio and we proved that
this scenario also leads to the same slow-roll epoch that the
$\beta(\phi)\sim 0$ condition produces.

Before  getting to the core of our analysis, let us mention that
in this article  we shall use the geometric background of a flat
Friedmann-Robertson-Walker (FRW) Universe, with line element,
\begin{equation}\label{frw}
ds^2 = - dt^2 + a(t)^2 \sum_{i=1,2,3} \left(dx^i\right)^2\, ,
\end{equation}
where $a(t)$  is the scale factor, and  $H=\frac{\dot{a}}{a}$ is
the Hubble rate.

\section{Brief Overview of the TCC Problem of Slow-roll Inflation and How a Power-law Tail Can Resolve this Problem}

The TCC problem in scalar field theory indicates that the scale of
inflation should be $V_e<10^{10}\,$GeV, a fact that constrains the
tensor-to-scalar ratio to take quite small values, namely
$r<10^{-30}$. Now this TCC issue basically excludes most viable
cosmological scenarios in single scalar field inflation. However,
if one combines a power-law tail that follows a slow-roll era, the
TCC problems of slow-roll inflation find a consistent remedy. In
Ref. \cite{Odintsov:2025zrp} we provided such a theoretical
framework, for which a standard $R^2$ gravity era is followed by a
power-law inflationary era realized by a single scalar field. In
this work we aim to realize this slow-roll to power-law inflation
transition in the context of single scalar field theory. This
transition scheme from a slow-roll era to a power-law era seems to
provide a viable and consistent remedy to the TCC problems of
slow-roll inflation and let us see why. First let us note that we
will realize a single scalar field theory slow-roll era, followed
by a power-law era realized again by the same single scalar field.
The power-law era must have a quintessential equation of state and
the scalar potential during this era must be a strictly
exponential one, of the form,
\begin{equation}\label{potentialapprox}
V=V_0e^{\sqrt{\frac{6\beta}{\beta+2}}\kappa \phi}\, ,
\end{equation}
and also the scalar field during this era satisfies,
\begin{equation}\label{eoscondition}
\dot{\phi}^2=\beta V(\phi)\, ,
\end{equation}
a condition which is responsible for the exponential form of the
potential during this era and also produces a constant equation of
state (EoS) parameter for the scalar field, of the form,
\begin{equation}\label{eosscalarfinal}
w_{\phi}=\frac{\beta-2}{\beta+2}\, .
\end{equation}
The scale factor during this era is $a(t)=a_0\,t^{\eta}$, with
$\eta=\frac{\beta +2}{3 \beta }$ and the TCC problems are solved
when $\beta<1$ so we shall choose $\beta=0.99$ as in our recent
work Ref. \cite{Odintsov:2025zrp}.

Now there are several scenarios which can be realized and the TCC
problems can be solved. For example the slow-roll era may be
shorter than 60 e-foldings because the flatness problems of
inflation may be solved during the power-law era. Indeed, the
scalar field energy density at the end of the slow-roll era and at
the beginning of the power-law era is
\cite{Kamali:2020drm,Odintsov:2025zrp},
\begin{equation}\label{b1}
\rho_{\phi}\sim a^{-\frac{3\beta}{1+\frac{\beta}{2}}}\, .
\end{equation}
Now, the spatial flatness issue requires that
\begin{equation}\label{b2}
\Omega_K<10^{-2}\frac{T_0T_{eq}}{T_R^2}\, ,
\end{equation}
with $T_0$, $T_{eq}$ and $T_R$ being the temperatures of the
Universe at present day, during the matter-radiation equality and
at the end of the power-law inflationary tail. Hence it is
apparent that the spatial flatness issue may be solved
synergistically by the slow-roll era and the power-law era, no
matter how short is the slow-roll era. We demand that the total
decrease of the curvature density during the power-law
inflationary tail is overall larger that the relative increase
after the power-law inflationary tail, hence
\cite{Odintsov:2025zrp,Kamali:2020drm},
\begin{equation}\label{b3}
\left(\frac{a_i}{a_R}\right)^{2-\frac{3\beta}{1+\frac{\beta}{2}}}<10^{-2}\frac{T_0T_{eq}}{T_R^2}\,
,
\end{equation}
Also, we must require that the current Hubble radius comoving
scale is larger compared to the Hubble length at the beginning of
the power-law  inflationary tail,  $H^{-1}(t_i)$,
\begin{equation}\label{b4}
H^{-1}(t_i)<H_0^{-1}\frac{T_0 a_i}{T_Ra_R}\, .
\end{equation}
Hence, in terms of the potential we have
\cite{Odintsov:2025zrp,Kamali:2020drm},
\begin{equation}\label{b8}
\frac{V_R}{V_i}<\frac{T_R^2}{T_0T_{eq}}\left(\frac{a_i}{a_R}
\right)^2\, ,
\end{equation}
with $V_i$ being the scalar potential value when the inflationary
power-law tail commences. In terms of the Friedmann equation we
finally arrive to the inequality
\cite{Odintsov:2025zrp,Kamali:2020drm},
\begin{equation}\label{b10}
\left( \frac{a_i}{a_R}
\right)^{2-\frac{3\beta}{\sqrt{1+\frac{\beta}{2}}}}>\frac{T_0T_{eq}}{T_R^2}\,
,
\end{equation}
with $a_i$ and $a_R$ being the values of the scale factor at the
beginning of the power-law  inflationary tail and at the end of
the power-law inflationary tail. If power-law tail has
$\beta\simeq 0$, which is basically a slow-roll tail, the
inequalities of Eq. (\ref{b10}) and Eq. (\ref{b3}) would be in
conflict. However, when $\beta$ takes values in the range
$0.5<\beta<1$ the two inequalities are compatible. In this work we
shall take, $\beta\sim 0.99$ which results to a quintessential
type inflationary power-law tail with $w_{\phi}\sim -0.337793$.
Also following \cite{Odintsov:2025zrp,Kamali:2020drm}, the TCC
problems are resolved if the following inequality holds true,
\begin{equation}\label{tr}
\frac{T_R}{M_p}<\left( 3/
g^*\right)^{(1-\tilde{\beta})/(6-4\tilde{\beta})}\times10^{-2/(6-4\tilde{\beta})}\left(\frac{T_0T_{eq}}{M_p}
\right)^{1/(6-4\tilde{\beta})}\, ,
\end{equation}
with $\tilde{\beta}=\frac{3\beta}{\beta+2}$. In our case,
$\beta=0.99$, so we get, $T_R<10^{-30}\times M_p$, which does not
affect the flatness issue, which is solved by the inflationary
power-law tail, see the inequality (\ref{b10}) and also recall
that a slow-roll era precedes the power-law inflationary tail. We
have also prepared a schematic explanation for the question why
the slow-roll era followed by a power-law inflationary tail solves
the TCC problems.
\begin{figure}[h]
\centering
\includegraphics[width=17pc]{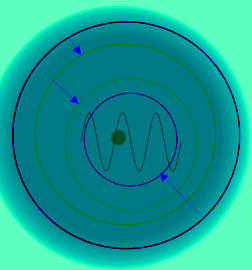}
\caption{The TCC problem resolution in standard slow-roll
inflation (blue circles) and in the scenario of a short slow-roll
era followed by a power-law inflationary tail (green circles).}
\label{plot1}
\end{figure}
The explanation is presented in Fig. \ref{plot1}. For a slow-roll
epoch, the Hubble horizon shrinks in an inverse exponential way,
thus the sub-Planckian modes are in risk in crossing the Hubble
horizon (blue circle). But if the slow-roll era is short and is
followed by a power-law inflationary tail (green circles), the
Hubble horizon shrinks with a smaller rate and thus more slowly
compared to the inverse exponential shrinking of the slow-roll
era. Thus the sub-Planckian modes always remain inside the Hubble
horizon during inflation.

\section{Smooth Slow-roll to Power-law Inflation Realization in Single Scalar Field Theory}

In order to realize a smooth transition from a slow-roll era to a
power-law inflationary tail era, in the context of single scalar
field theory, one must assume that the kinetic energy of the
scalar field satisfies the following condition,
\begin{equation}\label{mainrelation}
\dot{\phi}^2=\beta (\phi)V(\phi)\, ,
\end{equation}
and appropriately choose the function $\beta(\phi)$ in order to
realize the slow-roll power-law transition. The field equations of
the scalar field theory for a FRW metric are,
\begin{align}\label{eqnsofmkotion}
& 3 H^2=\kappa^2\left( \frac{1}{2}\dot{\phi}^2+V(\phi)\right)\,
,\\ \notag & -2\dot{H}=\kappa^2\dot{\phi}^2\, ,
\end{align}
and
\begin{equation}\label{scalareqnofmotion}
\ddot{\phi}+3H\dot{\phi}+V'(\phi)=0\, .
\end{equation}
If the kinetic energy of the scalar field behaves as in Eq.
(\ref{mainrelation}), then the total EoS of the scalar field
behaves as follows,
\begin{equation}\label{slowroll}
w_{\phi}=\frac{\beta(\phi)-2}{\beta(\phi)+2}\, .
\end{equation}
Also, we assume that the scalar field potential is positive, in
order to avoid instabilities and ensure that an inflationary era
can be realized, instead of some bounce era in the case of a
negative scalar potential. Now, the condition (\ref{mainrelation})
restricts the scalar function $\beta (\phi)$ to be positive, that
is $\beta(\phi)>0$. From Eq. (\ref{mainrelation}) by
differentiating with respect to the cosmic time, one gets,
\begin{equation}\label{ddotphi}
\ddot{\phi}=\frac{\beta'(\phi)V(\phi)+\beta(\phi)V'(\phi)}{2}\, .
\end{equation}
Also the Friedmann equation in Eq. (\ref{eqnsofmkotion}) becomes,
\begin{equation}\label{updatedfriedmann}
H^2=\frac{\kappa^2V(\phi)}{6}\left(\beta(\phi)+2\right)\, ,
\end{equation}
and also by substituting Eq. (\ref{ddotphi}) in the Klein-Gordon
equation (\ref{scalareqnofmotion}) we get,
\begin{equation}\label{updatedkleingordon}
H\dot{\phi}=-\frac{\beta'(\phi)V(\phi)+\beta(\phi)V'(\phi)+2V'(\phi)}{6}\,
.
\end{equation}
Taking the square of Eq. (\ref{updatedkleingordon}) and also using
Eq. (\ref{updatedfriedmann}) we finally get,
\begin{equation}\label{maindifferentialequation}
6\,\beta(\phi)V(\phi)^2\left(\beta(\phi)+2
\right)=\left(\beta'(\phi)V(\phi)+\beta(\phi)V'(\phi)+2V'(\phi)^2\,
. \right)
\end{equation}
Now this is a constraint equation between the scalar field
potential and the function $\beta(\phi)$ which essentially
determines the kinetic energy of the scalar field.

In order to realize the two distinct regimes, namely the slow-roll
regime and the power-law tail, one essentially needs primordially
$\beta(\phi)\simeq 0$ and also at a later point, the power-law
inflationary tail can be realized if Eq. (\ref{eoscondition}) is
realized, namely $\dot{\phi}^2=\beta V(\phi)$, so
$\beta(\phi)=\beta$ and $\beta $ must be chosen to be $\beta=0.99$
in order for the scenario of the previous section to be realized
and thus solve the TCC problems of slow-roll inflation. A choice
for which such a transition can be realized is the following,
\begin{equation}\label{choiceofbeta}
\beta(\phi)=\frac{\gamma}{\delta+\lambda \kappa \phi}\, ,
\end{equation}
where $\gamma$, $\delta$ and $\lambda$ are dimensionless
parameters. Now, the slow-roll era can be realized primordially
when the scalar field values are large and the following is
satisfied, $\lambda \kappa \phi\gg \delta ,\gamma$, thus when
$\lambda \kappa \phi \gg 1$ primordially, one has approximately,
\begin{equation}\label{apprxslow}
\beta(\phi)\simeq 0\, ,
\end{equation}
and when the scalar field takes smaller values at the end of the
slow-roll era, the approximation $\lambda \kappa \phi \gg 1$
breaks and thus when $\delta ,\gamma\gg \lambda \kappa \phi$ one
has approximately,
\begin{equation}\label{apprxslow1}
\beta(\phi)\simeq \frac{\gamma}{\delta}\, ,
\end{equation}
hence in order to have the desired power-law behavior of the
previous section one must choose the values of the parameters
$\gamma$ and $\delta$ as $\frac{\gamma}{\delta}=\beta=0.99$. We
shall use this constraint in the following. Now let us use the
functional form of $\beta(\phi)$ given in Eq. (\ref{choiceofbeta})
and we shall solve the differential equation
(\ref{maindifferentialequation}) in order to find the scalar
potential $V(\phi)$, and the result is,
\begin{equation}\label{scalarpotentialfull}
V(\phi)=V_0\,\frac{(\delta +\kappa  \lambda  \phi ) e^{
\frac{\sqrt{6} \gamma ^{3/2}}{\lambda  \sqrt{\gamma +2 \delta +2
\kappa  \lambda  \phi }}+\frac{2 \sqrt{6} \sqrt{\gamma } \kappa
\phi }{\sqrt{\gamma +2 \delta +2 \kappa  \lambda  \phi }}+\frac{2
\sqrt{6} \sqrt{\gamma } \delta }{\lambda  \sqrt{\gamma +2 \delta
+2 \kappa  \lambda  \phi }}}}{\gamma +2 \delta +2 \kappa \lambda
\phi }\, .
\end{equation}
Primordially, when the condition $\lambda \kappa \phi\gg \delta
,\gamma$ holds true, the potential can be approximated as follows,
\begin{equation}\label{approximatepotential1}
V(\phi)\simeq\mathcal{V}_0 \, e^{2 \sqrt{\frac{3 \gamma  \kappa
\phi }{\lambda }}}\, ,
\end{equation}
where $\mathcal{V}_0=\frac{V_0}{2}$. Now we can check the
phenomenology of the scalar potential
(\ref{approximatepotential1}) during the slow-roll era. The
slow-roll indices for the scalar theory during the slow-roll era
are,
\begin{equation}\label{slowrollindices1}
\epsilon=\frac{1}{2 \kappa ^2}\left(\frac{V'(\phi )}{V(\phi
)}\right)^2\, ,
\end{equation}
\begin{equation}\label{slowrollindices2}
\eta=\frac{1}{\kappa ^2}\frac{V''(\phi )}{V(\phi )}\, .
\end{equation}
By solving the equation $\epsilon=\mathcal{O}(1)$ we get the value
of the scalar field at the end of the slow-roll era, which is,
\begin{equation}\label{phifinal}
\phi_f=\frac{3\gamma }{2 \kappa  \lambda }\, ,
\end{equation}
and also by using the $e$-foldings number,
\begin{equation}\label{efoldings}
N=\kappa^2\int_{\phi_f}^{\phi_i}\frac{V(\phi)}{V'(\phi)}\mathrm{d}\phi
\, ,
\end{equation}
by solving the resulting equation with respect to $\phi_i$ which
is the value of the scalar field at the beginning of the slow-roll
era, that is at first horizon crossing, we get,
\begin{equation}\label{phiinitial}
\phi_i=\frac{3 \sqrt[3]{\frac{\gamma ^3}{\lambda ^3}+\frac{2
\sqrt{2} \sqrt{\gamma } \gamma ^{3/2} N}{\lambda ^2}+\frac{2
\gamma  N^2}{\lambda }}}{2 \kappa }\, .
\end{equation}
Having the value of the scalar field at first horizon crossing, we
can evaluate the spectral index of the primordial scalar
perturbations and the corresponding tensor-to-scalar ratio at
first horizon crossing and thus determine when the inflationary
phenomenology is viable. The spectral index $n_{s}$ and the
tensor-to-scalar ratio in terms of the slow-roll indices are,
\begin{equation}\label{spectralindexslow}
n_s=1-6\epsilon+2\eta,\,\,\,r=16\,\epsilon \, .
\end{equation}
Now we can investigate the parameter space of the above model and
determine the values of the free parameters that render the
slow-roll inflationary regime a viable one. After some
investigation we have found some interesting results which in a
nutshell require a short slow-roll era, with about 30 $e$-foldings
and the parameter $\lambda$ must be in the range $\lambda \sim
\mathcal{O}(10-100)$, while the parameter $\gamma$ must be of the
order $\gamma \sim \mathcal{O}(0.1)$. Before presenting the full
analysis of the parameter space and the confrontation of the
inflationary phenomenology with the Planck 2018
\cite{Planck:2018jri}, the latest ACT data \cite{ACT:2025tim} and
the updated BICEP-Keck constraints on the tensor-to-scalar ratio
\cite{BICEP:2021xfz}, let us investigate whether the approximation
$\lambda \kappa \phi\gg \delta ,\gamma$ holds true at the
beginning of the slow-roll era and also investigate the behavior
of the scalar field when it takes values near $\phi\sim \phi_f$.
By taking $\lambda \sim \mathcal{O}(100)$, $\gamma \sim
\mathcal{O}(0.1)$ for 30 $e$-foldings we have $\phi_i\sim
1.82\,M_p$ and also $\phi_f\sim 0.0015\,M_p$ so $\lambda \kappa
\phi_i\sim 364$ and $\lambda \kappa \phi_f\sim 0.3$. Hence the
constraint $\lambda \kappa \phi\gg \delta ,\gamma$ is well
respected at the beginning of the inflationary era and until the
very end of the inflationary era, where $\lambda \kappa \phi_f
\sim \gamma+2\delta$. Now an important feature of this scenario is
that the slow-roll era lasts for nearly 30 $e$-foldings, which is
a rather short inflationary period. However recall that this
slow-roll era will be followed by another inflationary era of
power-law type, therefore the flatness issues which require a long
inflationary period, are solved due to the power-law inflationary
era that follows the initial slow-roll regime. In fact this type
of behavior is rather appealing since if the original slow-roll
era lasted for too long, this could be problematic because the
combination of the slow-roll and the power-law inflationary eras
could produce an overall very long inflationary era.

\begin{figure}[h]
\centering
\includegraphics[width=35pc]{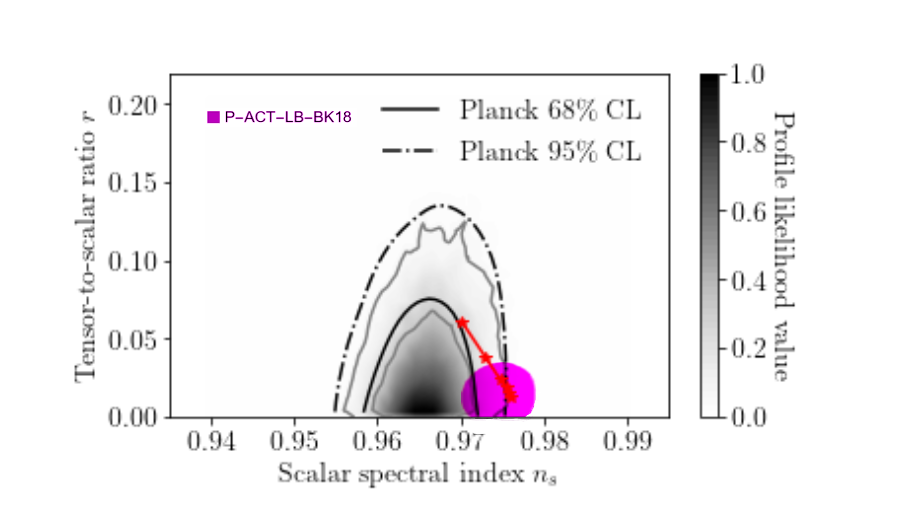}
\caption{The phenomenology of the slow-roll regime potential
(\ref{approximatepotential1}) for $\gamma=0.1$ and  $N=30$
$e$-foldings, with the parameter $\lambda$ chosen in the range
$\lambda=[10,100]$. We confront the model with the Planck 2018
likelihood curves, the BICEP/Keck updates on the Planck data and
the ACT constraints on the spectral index (purple region). The
model is compatible with the ACT data for $\lambda$ in the range
$\lambda=[40,100]$. } \label{plot2}
\end{figure}

Now let us analyze the phenomenology of the slow-roll regime, and
in Fig. \ref{plot2} we present the spectral index and
tensor-to-scalar ratio for $\gamma=0.1$ and  $N=30$ $e$-foldings,
with the parameter $\lambda$ taking values in the range
$\lambda=[10,100]$, confronted with the Planck 2018 curves and the
ACT data, and also with the updated constraints on the
tensor-to-scalar ratio \cite{BICEP:2021xfz}. It is worth quoting
the constraints here for convenience. The Planck 2018 data
constrain the tensor-to-scalar ratio and the spectral index to be,
\begin{gather}
n_{s}=0.9649\pm{}0.0042 , \hspace{0.3cm} r<0.064 \label{eq:31}
\end{gather}
while the BICEP/Keck updates on the Planck data
\cite{BICEP:2021xfz} further constrain the tensor-to-scalar ratio
to be,
\begin{equation}\label{tensortoscalarratio}
r<0.036\, ,
\end{equation}
at $95\%$ confidence. Finally, the recent ACT data constrain the
spectral index to be \cite{ACT:2025tim},
\begin{equation}\label{actspectral}
n_s=0.974\pm 0.003 \, .
\end{equation}
Now as it can be seen in Fig. \ref{plot2} the slow-roll regime of
the model with potential (\ref{approximatepotential1}) is
compatible with the ACT data \cite{ACT:2025tim} and with the
BICEP/Keck updates on the Planck data \cite{BICEP:2021xfz},
especially for values of the parameter $\lambda$ in the range
$\lambda=[40,100]$.
Now, let us comment on the issue that the initial value of the
scalar field is super-Planckian, but this feature often occurs in
scalar field theories, see for example \cite{Silverstein:2008sg},
since the scalar fields are the moduli low-energy remnants of
stringy origin.

Regarding the regime where the scalar field values are much
smaller than the Planck scale, the potential of Eq.
(\ref{scalarpotentialfull}) is approximated as,
\begin{equation}\label{regimepowerlaw}
V(\phi)\simeq \mathcal{A}_0\,e^{\frac{\sqrt{6} \sqrt{\gamma }
\kappa  \phi }{\sqrt{\gamma +2 \delta }}}\, ,
\end{equation}
where $\mathcal{A}_0$ is defined to be,
\begin{equation}\label{mathcala}
\mathcal{A}_0=V_0\,e^{\frac{\sqrt{6} \gamma ^{3/2}+2 \sqrt{6}
\sqrt{\gamma } \delta }{\lambda  \sqrt{\gamma +2 \delta }}}\, .
\end{equation}
Now it can easily be checked that the potential of Eq.
(\ref{regimepowerlaw}) is generated from the condition
$\dot{\phi}^2=\beta V(\phi)$ with $\beta=\frac{\gamma}{\delta}$.
This is a proof that the scenario with the potential
(\ref{scalarpotentialfull}) is self-consistent, since in the small
field values regime it can realize a power-law inflationary era
which naturally stems from the condition $\dot{\phi}^2=\beta
V(\phi)$ with the desired form of the parameter $\beta$. This is a
proof that this scenario is indeed self-consistent. We need to
note though that the amplitude of the potential during the
power-law era is different from the initial amplitude, and this
could be an issue since if the initial amplitude is compatible
with the Planck constraints, the amplitude (\ref{mathcala}) is
almost half of the original amplitude. However we need to stress
two things, firstly the regime for which the scalar potential is
described by Eq. (\ref{regimepowerlaw}) occurs when
$\gamma,\delta\gg \kappa \lambda \phi$, so there might be an
intermediate era with $\gamma,\delta\sim \kappa \lambda \phi$.
Secondly, the modes which exit the horizon during the power-law
inflationary regime do not necessarily contribute to the linear
CMB spectrum. This is because during the power-law inflationary
era, the Hubble horizon shrinks at a smaller rate, compared to a
de Sitter evolution, thus even though inflation occurs for some
large $e$-foldings number, the modes that enter the horizon during
this regime might have a small wavelength, equal or smaller to
10Mpc, thus these are not expected to contribute to the linear
part of the CMB probed by the experiments. It is notable though
that the amplitude of those perturbations are smaller compared to
the amplitude of the scalar perturbations during the slow-roll
regime.

Now, let us adopt a distinct line of research compared to the two
inflationary patches approach we analyzed so far. Let us consider
directly the inflationary regime caused by the condition
(\ref{mainrelation}) and use the full potential
(\ref{scalarpotentialfull}) instead of the slow-roll inflationary
patch of Eq. (\ref{approximatepotential1}). In order to analyze
this regime, we must calculate the first and second slow-roll
indices from first principles, instead of using $\epsilon$ and
$\eta$ defined in Eqs. (\ref{spectralindexslow}), which hold true
only if the slow-roll assumption holds true. So we need to
calculate the inflationary indices $\epsilon_1$ and $\epsilon_2$,
defined as follows \cite{reviews1},
\begin{equation}\label{slowrollindicesnew}
\epsilon_1=-\frac{\dot{H}}{H^2},\,\,\,\epsilon_2=\frac{\ddot{\phi}}{H\dot{\phi}}\,
.
\end{equation}
In terms of these slow-roll indices, the spectral index of the
scalar perturbations and the tensor-to-scalar ratio read
\cite{reviews1},
\begin{equation}\label{spectralindexnoslowroll}
n_s=1-4\epsilon_1-2\epsilon_2,\,\,\,r=16\epsilon_1\, .
\end{equation}
Now assuming that the condition (\ref{mainrelation}) holds true,
we will calculate the inflationary indices
(\ref{slowrollindicesnew}) directly from the Friedmann equation
(\ref{updatedfriedmann}) and the Raychaudhuri equation which in
view of Eq. (\ref{mainrelation}) it becomes,
\begin{equation}\label{raychaudhuri}
\dot{H}=-\frac{\kappa^2\beta(\phi)V(\phi)}{2}\, .
\end{equation}
In view of Eqs. (\ref{updatedfriedmann}), (\ref{raychaudhuri}),
(\ref{ddotphi}) and (\ref{updatedkleingordon}), the first
slow-roll index $\epsilon_1$ reads,
\begin{equation}\label{epsilon1finalform}
\epsilon_1=\frac{3\beta (\phi)}{\beta(\phi)+2}\, ,
\end{equation}
while $\epsilon_2$ takes the form,
\begin{equation}\label{epsilon2finalform}
\epsilon_2=-\frac{3\left(\beta'(\phi)V(\phi)+\beta(\phi)V'(\phi)\right)}{\beta'(\phi)V(\phi)+\beta(\phi)V'(\phi)+2V'(\phi)}\,
.
\end{equation}
Also, the $e$-foldings number defined as ,
\begin{equation}\label{efoldingsupdated}
N=\int_{\phi_i}^{\phi_f}\frac{H^2}{H\dot{\phi}}\mathrm{d}\phi\, ,
\end{equation}
takes the final form,
\begin{equation}\label{finalformefoldings}
N=\kappa^2\int_{\phi_f}^{\phi_i}\frac{V(\phi)\left(\beta(\phi)+2\right)}{\beta'(\phi)V(\phi)+\beta(\phi)V'(\phi)+2V'(\phi)}\mathrm{d}\phi\,
.
\end{equation}
Now we can proceed to examine the phenomenology of the complete
potential (\ref{scalarpotentialfull}) and our analysis verified
that the inflationary regimes generated by the approximate
potential (\ref{approximatepotential1}) is completely identical
with the phenomenology of the model (\ref{scalarpotentialfull})
for the same values of the free parameters and the $e$-foldings
number. This result is remarkable and validates our approach to
separate the inflationary era to two distinct patches, one
slow-roll and one power-law inflationary era, depending on the
values of the scalar field. In fact, even the scalar field values
are the same for the models (\ref{approximatepotential1}) and
(\ref{scalarpotentialfull}).

\section{The Power-law Era and its Duration}

Now let us consider the duration of the power-law era issue, which
is important in this smooth transition scheme we introduced in the
previous section. Notice that the slow-roll regime lasts for about
30 $e$-foldings, therefore it is important for the power-law era
to last for at least 30 more $e$-foldings. However, the power-law
regime generated by an exponential potential is known to lead to a
constant first slow-roll index. In our case, the first slow-roll
index is,
\begin{equation}\label{firstslowrollpowerlaw}
\epsilon_1=-\frac{\dot{H}}{H^2}=\frac{\beta +2}{3 \beta  t}\, ,
\end{equation}
since recall that $a(t)=a_0\,t^{\eta}$, in our case, with
$\eta=\frac{\beta +2}{3 \beta }$ and also recall that the TCC
problems are solved when $\beta<1$ and we choose $\beta=0.99$ as
in our recent work Ref. \cite{Odintsov:2025zrp}. Hence, the first
slow-roll index $\epsilon_1$ in our case is $\epsilon_1=0.993311$.
Therefore, it is apparent that the power-law regime in our case is
not a slow-roll regime at all. This makes an importance difference
regarding the evolution of the scalar field during this regime.
Let us highlight our thinking and for simplicity let us switch to
Planck units with $M_p=1$. Firstly let us use some notation, and
let us recall the potential (\ref{approximatepotential1}) during
the power-law regime which is $V(\phi)=\mathcal{V}_0\,
e^{\tilde{\lambda} \phi}$ where
$\tilde{\lambda}=\sqrt{\frac{6\beta}{\beta+2}}$. Now for this
potential we can obtain an analytical solution for the
Klein-Gordon equation for the flat FRW metric. Indeed, since the
scale factor is $a(t)=a_0\,t^{\eta}$, we get the following
solution,
\begin{equation}\label{eq}
\phi(t) = -\frac{2}{\tilde{\lambda}} \ln t + \text{const}\, ,
\end{equation}
and we can compute directly the number of $e$-foldings as follows,
\begin{equation}\label{dfe}
N = \ln \left( \frac{a(t_f)}{a(t_i)} \right) = \eta \ln\left(
\frac{t_f}{t_i} \right)\, .
\end{equation}
Relating time to the scalar field, we invert,
\begin{equation}\label{defr}
\phi(t) = -\frac{2}{\tilde{\lambda}} \ln t + \text{const} \quad
\Rightarrow \quad t \propto e^{-\lambda \phi / 2}
\end{equation}
so we get,
\begin{equation}\label{dfeerd}
N = \eta \ln \left( \frac{t_f}{t_i} \right) =
-\frac{2}{\tilde{\lambda}^2} \cdot \frac{\tilde{\lambda}}{2}
(\phi_f - \phi_i) = \frac{1}{\tilde{\lambda}} (\phi_f - \phi_i)\,
,
\end{equation}
where in this case $t_f$ and $t_i$ are the cosmic instances that
the power-law inflationary regime starts and ends, and $\phi_f$
and $\phi_i$ are the corresponding scalar field values, thus the
result is,
\begin{equation}\label{dfred}
N = \frac{1}{\tilde{\lambda}} (\phi_f - \phi_i)\, .
\end{equation}
Now the big difference between the non-slow-roll approach and the
slow-roll approach is that the scalar field in the non-slow-roll
approach actually increases as the time passes. This is in
contrast to the slow-roll approach, where the scalar field
degreases, and the $e$-foldings number takes the form,
\begin{equation}\label{slowrollfaof}
N = \int_{\phi_i}^{\phi_f} \frac{H}{\dot{\phi}} \, d\phi =
\int_{\phi_i}^{\phi_f}
\frac{\sqrt{V(\phi)/3}}{-\frac{\tilde{\lambda}}{\sqrt{3}}
\sqrt{V(\phi)}} \, d\phi = -\frac{1}{\tilde{\lambda}}
\int_{\phi_i}^{\phi_f} d\phi = \frac{1}{\tilde{\lambda}}(\phi_i -
\phi_f)\, .
\end{equation}
Now having in mind that the the scalar field values increase in
the non-slow-roll power-law inflation scenario we consider, we
need to find a way to end the inflationary regime in a concrete
way, since the power-law inflationary regime is indefinite. There
are two ways to end this regime, the first is to use a waterfall
scalar field which is purpose is to end the inflationary era via
some sort of phase transition occurring in its sector. Scalar
fields are remnants from the ultra-violet completion of classical
theories thus these are expected to exist in nature. Thus one way
to end the inflationary regime is to use a  hybrid inflationary
regime
\cite{Felder:2001kt,Linde:1993cn,Copeland:1994vg,Copeland:2000hn,Lee:2023dcy}
employing an auxiliary scalar $\chi$ and its couplings to the
inflaton scalar. The combined inflation-waterfall scalar potential
has the following form,
\begin{equation}\label{hybrid}
V(\phi, \chi) = \mathcal{V}_0 e^{\tilde{\lambda} \phi / M_{p}} -
\frac{1}{2} g^2 \phi^2 \chi^2 - \frac{\lambda_\chi}{4} \left(
\chi^2 - v^2 \right)^2,
\end{equation}
with $g$ being the coupling constant between the inflaton $\phi$
and the waterfall scalar field $\chi$, while $\lambda_\chi$ is the
self-coupling of the waterfall scalar, and $v$ will be the vacuum
expectation value of the waterfall scalar once its symmetry is
broken. The effective potential (\ref{hybrid}) indicates that
initially, for small values of the scalar field, at the beginning
of the power-law inflationary regime, the effective mass squared
of the waterfall field $\chi$ is,
\begin{equation} m_\chi^2(\phi)=\left(\frac{\partial^2 V(\phi, \chi)}{\partial \chi^2} \right)\Big{|}_{\chi=0} = \lambda_\chi v^2 - g^2 \phi^2
\, .
\end{equation}
During the power-law inflationary regime, the scalar field $\phi$
values increase, thus there exists a critical value of the scalar
field $\phi$, namely $\phi_c$, at which the effective mass of
$\chi$ gets equal to zero, that is, $m_\chi^2(\phi_c) = 0$, which
is,
\begin{equation}
\phi_c = \frac{\sqrt{\lambda_\chi}}{g} v.
\end{equation}
Hence, when $\phi < \phi_c$, the effective mass of the waterfall
scalar is positive $m_\chi^2> 0$, and $\phi > \phi_c$, then
$m_\chi^2 < 0$ and the $\chi=0$ minimum becomes unstable, thus the
waterfall scalar undergoes some rapid phase transition, first or
second order and the waterfall field acquires a new expectation
value $\chi = \pm v$. It is exactly this rapid phase transition
that ends the power-law inflationary regime. According to our our
previous findings, the total duration of the power-law
inflationary regime is,
\begin{equation} N =
\frac{1}{\tilde{\lambda}  M_{p}} (\phi_c - \phi_i).
\end{equation}
Therefore, the total duration of the power-law era clearly depends
on the specifics of the $\chi$ transition and also depends
strongly on the parameter $\tilde{\lambda}$ which is
$\tilde{\lambda}=\sqrt{\frac{6\beta}{\beta+2}}$. Also $\phi_i$ in
the above equation is basically the value of the scalar field when
the power-law inflationary regime starts, thus it is identical
with the value of the scalar field when the slow-roll era of the
previous section ends, thus it is give in Eq. (\ref{phifinal}),
therefore $\phi_i=\frac{3\gamma }{2 \kappa  \lambda }$. Hence,
since $\phi_c=\frac{\sqrt{\lambda_\chi}}{g} v$ by appropriately
choosing the parameters $v$ and $\lambda_\chi$ one can easily
achieve having a duration of the power-law inflationary regime of
the order $N\sim 30$.

\section{Conclusions}

In this article we studied how it is possible to realize a smooth
transition from a slow-roll to a power-law inflationary era within
the context of single scalar field theory. Our motivation is the
TCC problems of the standard slow-roll inflationary paradigm of
single scalar field theory. Specifically, the inflationary era is
considered to belong to the classical regime of our Universe,
since during the inflationary era the Hubble horizon shrinks and
the modes that exit the horizon freeze outside the horizon and are
thus classical modes.  Some of these modes become relevant for
present day observations, after they re-enter the horizon since
these modes constitute the CMB fluctuations. However, in the
standard slow-roll era, the Hubble horizon shrinks in an inverse
nearly quasi de-Sitter way, thus the horizon shrinks too fast,
hence it is possible that modes with very small wavelength exit
the horizon during the inflationary era. Therefore,
trans-Planckian modes may exit the horizon, hence modes of the
extreme quantum epoch may cross the horizon and thus these could
be part of the classical regime. In order to avoid this in the
standard slow-roll paradigm, one must constraint significantly the
scale of inflation and the tensor-to-scalar theory. In order to
avoid this, in Ref. \cite{Odintsov:2025zrp} we showed that if the
standard slow-roll era realized by an $R^2$ gravity is followed by
a power-law tail, then the TCC problems of standard inflation may
be resolved. Our work was inspired by the approach firstly
developed in \cite{Kamali:2020drm}. In this work we adopted the
same line of reasoning, and we investigated the way that a
slow-roll era may be smoothly followed by a power-law inflationary
era in the context of single scalar field inflation. As we showed,
by using a condition for the scalar field kinetic energy of the
form $\dot{\phi}^2=\beta (\phi)V(\phi)$, we may naturally realize
a smooth slow-roll to power-law inflationary era, if the function
$\beta (\phi)$ is suitably chosen. We studied the inflationary
regimes that can be generated in our approach using two distinct
formalisms, a direct approach by studying the slow-roll
inflationary regime for which $\dot{\phi}^2\simeq 0$ and an
indirect formalism in which we took into account that
$\dot{\phi}^2=\beta (\phi)V(\phi)$ holds true during the whole
inflationary regime. To this end, we had to calculate the
slow-roll indices and all the relevant quantities for inflationary
observables. Our results indicated that the two approaches yielded
identical results regarding the slow-roll regime, followed by a
power-law regime when the scalar field values drop significantly.
Thus our study indicated that a smooth transition from a slow-roll
to a power-law inflationary era can be achieved in the context of
single scalar field inflation. This transition ensures that the
power-law tail significantly delays the shrinking of the Hubble
horizon and therefore may provide a consistent remedy for the TCC
problems of standard slow-roll inflation.

\section*{Acknowledgements}

This work was partially supported by the program Unidad de
Excelencia Maria de Maeztu CEX2020-001058-M, Spain (S.D.O). This
research has been funded by the Committee of Science of the
Ministry of Education and Science of the Republic of Kazakhstan
(Grant No. AP26194585) (S.D. Odintsov and V.K. Oikonomou).

\end{document}